\RequirePackage{ifpdf}
\ifpdf 
\documentclass[pdftex]{sigma}
\else
\documentclass{sigma}
\fi

\begin{document}
\allowdisplaybreaks

\renewcommand{\PaperNumber}{018}

\FirstPageHeading

\ShortArticleName{Mathematical Analysis of a Generalized Chiral Quark Soliton Model}

\ArticleName{Mathematical Analysis of a Generalized Chiral Quark Soliton Model}

\Author{Asao ARAI}
\AuthorNameForHeading{A. Arai}

\Address{Department of Mathematics, Hokkaido University, Sapporo, 060-0810, Japan}
\Email{\href{mailto:arai@math.sci.hokudai.ac.jp}{arai@math.sci.hokudai.ac.jp}}

\ArticleDates{Received October 18, 2005, in f\/inal form January 25, 2006; Published online February 03, 2006}

\Abstract{A generalized version of the so-called chiral quark soliton model (CQSM)
in nuclear physics is introduced. The Hamiltonian of the generalized CQSM
is given by a Dirac type operator with a mass term being an operator-valued function. Some mathematically rigorous  results on the model are reported.
The subjects included are: (i)  supersymmetric structure; (ii)
spectral properties; (iii) symmetry reduction; (iv) a unitarily equivalent model.}


\Classification{81Q10; 81Q05; 81Q60; 47N50}

\section{Introduction}

The chiral quark soliton model (CQSM) \cite{Sawado}
is  a model describing a low-energy ef\/fective theory of the
quantum chromodynamics, which was developed in 1980's
(for physical aspects of the model, see, e.g., \cite{Sawado} and references
therein). The Hamiltonian of the CQSM
 is given by a Dirac type operator with iso-spin, which
dif\/fers from the usual Dirac type operator
in that the mass term is a matrix-valued function
with an ef\/fect of an interaction between quarks and the pion f\/ield.
It is an interesting object from the purely operator-theoretical
point of view too. But there are few
mathematically rigorous analyses
for such Dirac type operators (e.g., \cite{KY}, where
the problem on essential self-adjointness of
a Dirac operator with a variable mass term given by
a scalar  function is discussed).

In the previous paper \cite{AHS} we studied some fundamental
aspects of the CQSM in a mathematically rigorous way.
In this paper we  present a slightly general
form of the CQSM, which we call a {\it generalized CQSM},  and report that results similar to those
in \cite{AHS} hold on this model too, at least, as far as
some general aspects are concerned.

\section{A Generalized CQSM}

The Hilbert space of a Dirac particle with mass $M>0$
and iso-spin $1/2$ is taken to be
$L^2({\mathbb R}^3;{\mathbb C}^4)\otimes {\mathbb C}^2$.
For a generalization, we replace the iso-spin space
${\mathbb C}^2$ by an abitrary complex Hilbert space ${\cal K}$.
Thus the Hilbert space ${\cal H}$
in which we work  in the present paper is given by
\[
{\cal H}:=L^2({\mathbb R}^3;{\mathbb C}^4)\otimes {\cal K}.
\]

We denote by ${\sf B}({\cal K})$ the Banach space of
all bounded linear operators on ${\cal K}$ with domain ${\cal K}$.
Let
$T:{\mathbb R}^3 \to {\sf B}({\cal K})$; ${\mathbb R}^3\ni {\boldsymbol x}=(x_1,x_2,x_3)
\mapsto T({\boldsymbol x})
\in {\sf B}({\cal K})$ be a Borel measurable mapping
such that, for all ${\boldsymbol x}\in {\mathbb R}^3$,
 $T({\boldsymbol x})$ is a non-zero bounded self-adjoint operator
on ${\cal K}$ such that $\|T\|_{\infty}:=\sup\limits_{{\boldsymbol x}\in {\mathbb R}^3}\|T({\boldsymbol x})\| <\infty$, where
$\|T({\boldsymbol x})\|$ denotes the operator norm of $T({\boldsymbol x})$.

\begin{example}\label{ex1} In the original CQSM, ${\cal K}={\mathbb C}^2$
and $T({\boldsymbol x})={\boldsymbol \tau}\cdot {\boldsymbol n}({\boldsymbol x})$, where
${\boldsymbol n}:{\mathbb R}^3 \to {\mathbb R}^3$ is a measurable
vector f\/ield with $|{\boldsymbol n}({\boldsymbol x})|=1$, a.e. (almost everywhere) ${\boldsymbol x}\in
{\mathbb R}^3$ and ${\boldsymbol \tau}=(\tau_1,\tau_2,\tau_3)$
is the set of the Pauli matrices.
\end{example}

We denote by $\{\alpha_1,\alpha_2,\alpha_3, \beta\}$
the Dirac matrices, i.e., $4\times 4$-Hermitian
matrices satisfying
\[
\{\alpha_j,\alpha_k\}=2\delta_{jk}, \qquad \{\alpha_j,\beta\}=0,
\qquad \beta^2=1, \qquad j,k=1,2,3,
\]
where $\{A,B\}:=AB+BA$.

Let $F:{\mathbb R}^3 \to {\mathbb R}$ be measurable, a.e., f\/inite  and
\[
U_F:=(\cos F)\otimes I
+i(\sin F)\gamma_5\otimes T,
\]
where $I$ denotes identity and  $\gamma_5:=-i\alpha_1\alpha_2\alpha_3$.
We set ${\boldsymbol \alpha}:=(\alpha_1,\alpha_2,\alpha_3)$ and
$\nabla:=(D_1,D_2,D_3)$ with $D_j$ being the generalized partial
dif\/ferential operator in the variable $x_j$.
Then the one particle Hamiltonian of a generalized CQSM
is def\/ined by
\[
H:=-i{\boldsymbol \alpha} \cdot \nabla \otimes I+M(\beta\otimes I)U_F
\]
acting in the Hilbert space ${\cal H}$.
For a linear operator $L$, we denote its domain by $D(L)$.
It is well-known that $-i{\boldsymbol \alpha} \cdot \nabla$
is self-adjoint with $D(-i{\boldsymbol \alpha} \cdot \nabla)=
\cap_{j=1}^3D(D_j)$.
Since the operator $M(\beta\otimes I)U_F$ is bounded and self-adjoint,
it follows that $H$ is self-adjoint with domain
$D(H)=\cap_{j=1}^3D(D_j\otimes I)
=H^1({\mathbb R}^3;{\mathbb C}^4\otimes {\cal K})$, the
Sobolev space of order $1$ consisting of
${\mathbb C}^4\otimes {\cal K}$-valued measurable functions on ${\mathbb R}^3$.
In the context of the CQSM, the function $F$ is called a prof\/ile
function.
In what follows we sometimes omit
the symbol of tensor product $\otimes$ in writing equations down.

\begin{example} Usually
prof\/ile functions
are assumed to be
rotation invariant with boundary conditions
\[
F(0)=-\pi, \qquad \lim_{|{\boldsymbol x}|\to\infty}F({\boldsymbol x})=0.
\]
The following are concrete examples \cite{Sawado2}:
\begin{alignat*}{3}
& {\rm (I)} && F({\boldsymbol x})=-\pi \exp(-|{\boldsymbol x}|/R), \qquad R=0.55\times 10^{-15}\,{\rm m};&\\
& {\rm (II)} && F({\boldsymbol x})=-\pi \{a_1\exp(-|{\boldsymbol x}|/R_1)+a_2\exp(-|{\boldsymbol x}|^2/R_2^2)\},&\\
&&& a_1=0.65, \qquad R_1=0.58 \times 10^{-15}\,{\rm m}, \qquad a_2=0.35, \qquad R_2=\sqrt{0.3} \times 10^{-15}\,{\rm m};&\\
& {\rm (III)}\ \  && F({\boldsymbol x})=-\pi
\left(1-\frac{|{\boldsymbol x}|}{\sqrt{\lambda^2+|{\boldsymbol x}|^2}}\right),\qquad
\lambda=\sqrt{0.4}\times 10^{-15}\,{\rm m}.
\end{alignat*}
\end{example}

We say that a self-adjoint operator $A$ on ${\cal H}$
has chiral symmetry if $\gamma_5A \subset A\gamma_5$.

\begin{proposition} The Hamiltonian  $H$ has no chiral symmetry.
\end{proposition}

\begin{proof}
It is easy to check that, for all $\psi \in D(H)$,
$\gamma_5\psi \in D(H)$ and
$[\gamma_5,H]\psi=2M\gamma_5\beta U_F\psi$.
Note that $U_F\not=0$. Hence, $[\gamma_5,H]\not=0$ on $D(H)$.
\end{proof}

We note that, if $F$ and $T$ are dif\/ferentiable on ${\mathbb R}^3$ with $
\!\sup\limits_{{\boldsymbol x}\in {\mathbb R}^3} \!\! |\partial_jF({\boldsymbol x})|\! <\!\infty$ and
$\!\sup\limits_{{\boldsymbol x}\in {\mathbb R}^3}\!\!\|\partial_jT({\boldsymbol x})\|\!<\!\infty$  ($j=1,2,3$),
 then the square of $H$ takes the form
\[
H^2=(-\Delta +M^2)\otimes I-i
M\beta {\boldsymbol \alpha}\cdot(\nabla U_F)
+M^2\sin^2F\otimes (T^2-I).
\]
This is  a Schr\"odinger operator with an operator-valued potential.

\section{Operator matrix representation}

For more detailed  analyses of the model,
it is convenient to work with a suitable
representation of the Dirac matrices.
Here we
take the following representation of $\alpha_j$ and $\beta$
(the Weyl representation):
\[
\alpha_j=\left(
\begin{array}{cc}
\sigma_j & 0 \\
0 & -\sigma_j
\end{array}
\right), \quad
\beta=\left(
\begin{array}{cc}
0 & 1 \\
1 & 0
\end{array}
\right),
\]
where $\sigma_1$, $\sigma_2$ and $\sigma_3$ are the Pauli
matrices. Let
${\boldsymbol \sigma}:=(\sigma_1,\sigma_2,\sigma_3)$ and
\[
\Phi_F:=(\cos F)\otimes I+i(\sin F)\otimes T.
\]
Then  we have the following
operator matrix representation for $H$:
\[
H=\left(
\begin{array}{cc}
 -i{\boldsymbol \sigma}\cdot\nabla  &  M\Phi_F^*\vspace{2mm}\\
 M\Phi_F  &  i{\boldsymbol \sigma}
\cdot \nabla
\end{array}
\right).
\]

\section{Supersymmetric aspects}

Let $\xi:{\mathbb R}^3 \to {\sf B}({\cal K})$ be measurable such that,
for all ${\boldsymbol x}\in {\mathbb R}^3$,
$\xi({\boldsymbol x})$ is a bounded self-adjoint operator on ${\cal K}$
and $
\xi({\boldsymbol x})^2=I$, $\forall \; {\boldsymbol x}\in {\mathbb R}^3$.
Let
\[
\Gamma({\boldsymbol x}):=i\gamma_5\beta \otimes \xi({\boldsymbol x}), \quad {\boldsymbol x}\in {\mathbb R}^3.
\]
We def\/ine an operator $\hat \Gamma$ on ${\cal H}$ by
\[
(\hat \Gamma \psi)({\boldsymbol x}):=\Gamma({\boldsymbol x})\psi({\boldsymbol x}), \qquad \psi
\in {\cal H}, \qquad  {\rm a.e.}\ {\boldsymbol x}\in {\mathbb R}^3.
\]
The following fact is easily proven:

\begin{lemma} The operator $\hat \Gamma$ is self-adjoint and unitary,
i.e., it is a grading operator on ${\cal H}$:
$\hat \Gamma^*=\hat \Gamma$, $\hat \Gamma^2=I$.
\end{lemma}

\begin{theorem}\label{thm1} Suppose that
$\xi$ is strongly differentiable  with
$\sup\limits_{{\boldsymbol x}\in{\mathbb R}^3}
\|\partial_j\xi({\boldsymbol x})\|<\infty$ $(j=1,2,3)$  and
\begin{gather}
\sum_{j=1}^3\alpha_j \otimes D_j\xi({\boldsymbol x})
=M\gamma_5\beta\{\xi({\boldsymbol x}),T({\boldsymbol x})\}\sin F({\boldsymbol x}). \label{ST}
\end{gather}
Then $\hat \Gamma D(H)\subset D(H)$ and
$\{\hat \Gamma, H\}\psi=0$, $\forall\; \psi\in D(H)$.
\end{theorem}

\begin{proof} For all $\psi\in D_0:=C_0^{\infty}({\mathbb R}^3)\otimes_{\rm alg}
({\mathbb C}^4\otimes {\cal K})$ ($\otimes_{\rm alg}$ denotes algebraic tensor product),
we have
\begin{gather}
D_j\hat\Gamma\psi=i\gamma_5\beta\otimes (D_j\xi)\psi+i\gamma_5\beta\otimes
\xi(D_j\psi).\label{Dj}
\end{gather}
By a limiting argument using the fact that $D_0$ is a core of $D_j\otimes I$,
we can show that, for all $\psi \in D(D_j)$,
$\hat \Gamma\psi$ is in $D(D_j)$ and
(\ref{Dj}) holds.
Hence, for all $\psi \in D(H)$, $\hat\Gamma\psi\in D(H)$ and
(\ref{Dj}) holds. Thus we have for all $\psi \in D(H)$
$\{\hat \Gamma, H\}\psi=C_1\psi+C_2\psi$ with
$C_1:=\sum\limits_{j=1}^3\{\gamma_5\beta\otimes\xi,\alpha_jD_j\}$ and
$C_2:=iM\{\gamma_5\beta\otimes \xi, \beta U_F\}$.
Using the fact that $\{\gamma_5,\beta\}=0$ and $[\gamma_5,\alpha_j]=0$
($j=1,2,3$), we obtain
$C_1\psi=-\gamma_5\beta(\sum\limits_{j=1}^3\alpha_jD_j\xi)\psi$.
Similarly direct computations yield $(C_2\psi)({\boldsymbol x})=-M\sin F({\boldsymbol x})
\otimes\{\xi({\boldsymbol x}),T({\boldsymbol x})\}\psi(x)$. Thus
(\ref{ST}) implies $\{\hat \Gamma,H\}\psi=0$.
\end{proof}

Theorem~\ref{thm1} means that, under its assumption,
$H$ may be interpreted as a generator of a~supersymmetry
with respect to $\hat\Gamma$.

\begin{example} Consider the case ${\cal K}={\mathbb C}^2$.
Let $f,g:{\mathbb R}^3\to {\mathbb R}$ be a continuously dif\/ferentiable
function such that
\[
\big(1+C^2\big)f({\boldsymbol x})^2+g({\boldsymbol x})^2=1.
\]
with a real constant $C\not=0$ and $
{\boldsymbol n}({\boldsymbol x}):=(f({\boldsymbol x}), Cf({\boldsymbol x}), g({\boldsymbol x}))$.
Then $|{\boldsymbol n}({\boldsymbol x})|=1$, $\forall\; {\boldsymbol x}\in {\mathbb R}^3$. Let
\[
\xi:=\frac {C}{\sqrt{1+C^2}}\tau_1-
\frac 1{\sqrt{1+C^2}}\tau_2,\qquad
T({\boldsymbol x}):={\boldsymbol \tau}\cdot {\boldsymbol n}({\boldsymbol x}).
\]
Then $\xi^2=I$ and $(\xi,T)$ satisf\/ies (\ref{ST}).
\end{example}

To state spectral properties of $H$, we recall
some def\/initions.
For a self-adjoint operator~$S$, we denote
by  $\sigma(S)$ the spectrum of $S$.
The point spectrum of $S$, i.e.,
the set of  all the eigenvalues of $S$ is denoted $\sigma_{\rm p}(S)$.
An isolated eigenvalue of $S$ with f\/inite multiplicity
is called a discrete eigenvalue of $S$. We denote by
$\sigma_{\rm d}(S)$
the set of all the discrete eigenvalues of $S$.
The set $\sigma_{\rm ess}(S):=\sigma(S)\setminus \sigma_{\rm d}(S)$
is called the essential spectrum of $S$.

\begin{theorem}\label{spec1}
Under the same assumption as in  Theorem~{\rm \ref{thm1}}, the
following holds:
\begin{enumerate}
\itemsep=0pt
\item[\rm (i)] $\sigma(H)$ is symmetric with respect to the origin
of ${\mathbb R}$, i.e.,
if $\lambda \in \sigma(H)$, then $-\lambda \in \sigma (H)$.
\item[\rm (ii)] $\sigma_{\rm \#}(H)$  ($\#={\rm p, d}$)
is symmetric with respect to the origin
of ${\mathbb R}$ with
\[
\dim\ker(H-\lambda)=\dim\ker(H-(-\lambda))
\]
for all $\lambda \in \sigma_{\rm \#}(H)$.
\item[\rm (iii)] $\sigma_{\rm ess}(H)$ is symmetric with respect to the origin of ${\mathbb R}$.
\end{enumerate}

\end{theorem}

\begin{proof} Theorem \ref{thm1} implies a
unitary equivalence of $H$ and $-H$ ($\hat \Gamma H\hat \Gamma^{-1}=-H$).
Thus the desired results follow.
\end{proof}

\begin{remark} Suppose  that the assumption of
Theorem \ref{thm1} holds.
In view of supersymmetry breaking, it is interesting
to compute $\dim \ker H$.
This is related to the index problem:
Let
\[
{\cal H}_+:=\ker(\hat\Gamma-1), \quad {\cal H}_-:=
\ker(\hat \Gamma+1)
\]
and
\[
H_{\pm}:=H|{\cal H}_{\pm}.
\]
Then $H_+$ (resp. $H_-$) is a densely def\/ined closed linear
operator from ${\cal H}_+$ (resp. ${\cal H}_-$)  to
${\cal H}_-$ (resp. ${\cal H}_+$) with
$D(H_+)=D(H)\cap {\cal H}_+$ (resp. $D(H_-)=D(H)\cap D(H_-)$). Obviously
\[
\ker H=\ker H_+\oplus \ker H_-.
\]
The analytical index of $H_+$ is def\/ined by
\[
{\rm index}(H_+):=\dim \ker H_+ -\dim\ker H_+^*,
\]
provided that at least one of $\dim
\ker H_+$ and $\dim \ker H_+^*$ is f\/inite.
We conjecture that, for a class of $F$ and $T$,
${\rm index}(H_+)=0$.
\end{remark}

\section[The essential spectrum and finiteness
of the discrete spectrum of $H$]{The essential spectrum and f\/initeness\\
of the discrete spectrum of $\boldsymbol{H}$}

\subsection[Structure of the spectrum of $H$]{Structure of the spectrum of $\boldsymbol{H}$}

\begin{theorem}\label{espec} Suppose that
$\dim {\cal K} <\infty$ and
\begin{gather}
\lim_{|{\boldsymbol x}| \to \infty}F({{\boldsymbol x}})=0. \label{F0}
\end{gather}
Then
\begin{gather}
\sigma_{\rm ess}(H)=(-\infty, -M]\cup [M,\infty), \label{ess1}\\
\sigma_{\rm d}(H)\subset (-M,M).\label{ess2}
\end{gather}
\end{theorem}

\begin{proof} We can rewrite $H$ as
$H=H_0\otimes I+V$ with
$H_0:=-i{\boldsymbol \alpha}\cdot\nabla+M\beta$ and
$V:=M(\beta\otimes I)$ $(U_F-I)$.
We denote by $\chi_{R}$ ($R>0$)  the characteristic
function of the set $\{{\boldsymbol x}\in {\mathbb R}^3
|\,|{\boldsymbol x}|<R\}$. It is well-known that, for all $z\in {\mathbb C}\setminus {\mathbb R}$,
$(H_0-z)^{-1}\chi_R$ is compact \cite[Lemma~4.6]{Thaller}.
Since ${\cal K}$ is f\/inite dimensional, it follows that
$(H_0\otimes I-z)^{-1}\chi_R\otimes I$ is compact. We have
\begin{gather*}
\|V({\boldsymbol x})\| \leq  M(|\cos F({\boldsymbol x})-1|+|\sin F({\boldsymbol x})|\|T\|_{\infty})
\leq  M\left(\frac {|F({\boldsymbol x})|^2}2+|F({\boldsymbol x})|\|T\|_{\infty}\right).
\end{gather*}
Hence, by (\ref{F0}), we have
$\lim\limits_{R\to \infty}\sup\limits_{|{\boldsymbol x}|>R}\|V({\boldsymbol x})\|=0$.
Then, in the same way as in the method described
on \cite[pp.~115--117]{Thaller}, we can show that, for all $z\in {\mathbb C}\setminus{\mathbb R}$,
$(H-z)^{-1}-(H_0\otimes I-z)^{-1}$   is compact.
Hence, by a general theorem
(e.g., \cite[Theorem~4.5]{Thaller}),
 $\sigma_{\rm ess}(H)=\sigma_{\rm ess}(H_0\otimes I)$.
 Since $\sigma_{\rm ess}(H_0)=(-\infty,-M]\cup [M,\infty)$
(\cite[Theorem~1.1]{Thaller}), we obtain
(\ref{ess1}). Relation (\ref{ess2}) follows from (\ref{ess1}) and
$\sigma_{\rm d}(H)=\sigma(H)\setminus \sigma_{\rm ess}(H)$.
\end{proof}

\subsection[Bound for the number of
discrete eigenvalues of $H$]{Bound for the number of
discrete eigenvalues of $\boldsymbol{H}$}

Suppose that $\dim{\cal K}<\infty$ and  (\ref{F0}) holds.
Then, by Theorem~\ref{espec},
 we can def\/ine the number
of discrete eigenvalues of $H$  counting multiplicities:
\begin{gather}
N_H:=\dim {\rm Ran}\, E_H((-M,M)),
\end{gather}
where $E_H$ is the spectral measure of $H$.

To estimate an upper bound
for $N_H$, we introduce a hypothesis
for $F$ and $T$:

\medskip

\noindent
{\bf Hypothesis (A).}
\begin{enumerate}\vspace{-2mm}\itemsep=0pt
\item[(i)]$T({\boldsymbol x})^2=I$, $\forall\; {\boldsymbol x}\in {\mathbb R}^3$ and
$T$ is strongly dif\/ferentiable  with
$\sum\limits_{j=1}^3(D_jT({\boldsymbol x}))^2$ being a multiplication operator by a
scalar function on ${\mathbb R}^3$.
\item[(ii)] $F \in C^1({\mathbb R}^3)$.
\item[(iii)] $\sup\limits_{{\boldsymbol x}\in {\mathbb R}^3}
|D_jF({\boldsymbol x})| <\infty$,  $\sup\limits_{{\boldsymbol x}\in{\mathbb R}^3}\|D_jT({\boldsymbol x})\|
<\infty$ ($j=1,2,3$).
\end{enumerate}

Under this assumption, we can def\/ine
\[
V_F({{\boldsymbol x}}):=\sqrt{
|\nabla F({{\boldsymbol x}})|^2+\sum_{j=1}^3(D_jT({\boldsymbol x}))^2
\sin^2F({{\boldsymbol x}})}.
\]

\begin{theorem}\label{NH} Let
$\dim {\cal K} <\infty$. Assume \eqref{F0} and Hypothesis~{\rm (A)}.
Suppose that
\[
C_F:=
\int_{{\mathbb R}^6}\frac{V_F({{\boldsymbol x}})
V_F({\boldsymbol y})}{|{{\boldsymbol x}}-{\boldsymbol y}|^2}\,
d{{\boldsymbol x}}d{\boldsymbol y} < \infty.
\]
Then $N_H$ is finite with
\[
N_H \leq \frac
{(\dim {\cal K})M^2C_F}{4\pi^2}.
\]
\end{theorem}

A basic idea for the proof of
Theorem \ref{NH} is as follows.
Let
\[
L(F):=H^2-M^2.
\]
Then we have
\[
L(F)=-\Delta+M\left(
\begin{array}{cc}
0 & W_F^*\\
W_F & 0
\end{array}
\right)
\]
with $W_F:=i{\boldsymbol \sigma}\cdot \nabla\Phi_F$.
Note that
\[
W_F^*W_F=W_FW_F^*=V_F^2.
\]
Let
\[
L_0(F):=-\Delta-MV_F.
\]
For a self-adjoint operator $S$,
we introduce a set
\[
N_-(S):= \mbox{the number of negative eigenvalues of $S$ counting multiplicities}.
\]
The following is a key lemma:

\begin{lemma}
\begin{gather}
N_H\leq N_-(L(F))\leq N_-(L_0(F)).\label{NN}
\end{gather}
\end{lemma}

\begin{proof} For each
$\lambda \in \sigma_{\rm d}(H)\cap (-M,M)$, we have $\ker(H-\lambda)
\subset \ker(L(F)-E_{\lambda})$ with $E_{\lambda}=\lambda^2-M^2 <0$.
Hence the f\/irst inequality of (\ref{NN}) follows.
The second inequality of (\ref{NN}) can be proven in the same manner as
in the proof of \cite[Lemma 3.3]{AHS}, which uses the min-max principle.
\end{proof}

On the other hand, one has
\[
N_-(L_0(F))\leq \frac{(\dim {\cal K})M^2C_F}{4\pi^2}
\]
(the Birman--Schwinger bound \cite[Theorem XIII.10]{RS4}).
In this way we can prove Theorem
 \ref{NH}.

As a direct consequence of Theorem~\ref{NH}, we have the following
fact on the absence of discrete eigenvalues of $H$:

\begin{corollary}  Assume \eqref{F0} and Hypothesis {\rm (A)}.
Let $(\dim {\cal K})M^2C_F <4\pi^2$. Then
$\sigma_{\rm d}(H)=\varnothing$,
i.e., $H$ has no discrete eigenvalues.
\end{corollary}

\section{Existence of discrete ground states}

Let $A$ be  a self-adjoint operator on a Hilbert space  and
bounded from below. Then
\[
E_0(A):=\inf\sigma(A)
\]
is f\/inite. We say that  $A$ has a {\it ground state}
if $E_0(A)\in  \sigma_{\rm p}(A)$.
In this case, a non-zero vector
in $\ker(A-E_0(A))$  is called a {\it ground state of $A$}.
Also we say that $A$ has a discrete ground state
if  $E_0(A)\in \sigma_{\rm d}(A)$.

\begin{definition}{\rm Let
\[
E_0^+(H):=\inf\left[\sigma(H) \cap [0,\infty)\right], \qquad
E_0^-(H):=\sup\left[\sigma(H) \cap (-\infty,0]\right].
\]
\begin{enumerate}\itemsep=0pt
\item[\rm (i)]
If $E_0^+(H)$  is an eigenvalue of $H$, then
we say that {\it $H$ has a positive energy ground state}
and we call a non-zero vector
in $\ker(H-E_0^+(H))$  a {\it positive
energy ground state } of $H$.
\item[\rm (ii)] If $E_0^-(H)$  is an eigenvalue of $H$, then
we say that {\it $H$ has a negative energy ground state}
and we call a non-zero vector
in $\ker(H-E_0^-(H))$  a {\it negative
energy ground state} of $H$.
\item[\rm (iii)]
If  $E_0^+(H)$ (resp. $E_0^-(H)$) is a discrete eigenvalue of $H$,
then
we say that $H$ has a {\it discrete
 positive} (resp. {\it negative}) {\it energy ground state}.
\end{enumerate}
}
\end{definition}

\begin{remark} If the spectrum of $H$
is symmetric  with respect to
the origin of ${\mathbb R}$ as in Theorem~\ref{spec1},
then $E_0^+(H)=-E_0^-(H)$, and
$H$ has a positive energy ground state if and only if
it has a~negative energy ground state.
\end{remark}

Assume Hypothesis (A). Then the operators
\[
S_{\pm}(F):=-\Delta \pm M(D_3\cos F)
\]
are self-adjoint with $D(S_{\pm}(F))=D(\Delta)$ and bounded from below.

As for existence of discrete ground states of the Dirac
operator $H$, we have the following theorem:

\begin{theorem}\label{ex-gs} Let $\dim{\cal K}<\infty$.
Assume Hypothesis {\rm (A)} and \eqref{F0}.
Suppose that $E_0(S_+(F)) < 0$ or
$E_0(S_-(F)) <0$.
Then $H$ has
a discrete positive energy ground state or
a discrete negative ground state.
\end{theorem}

\begin{proof} We describe only an
 outline of proof. We have
\[
\sigma_{\rm ess}(L(F))=[0,\infty),
\qquad \sigma_{\rm d}(L(F))\subset [-M^2,0).
\]
Hence, if $L(F)$ has a discrete eigenvalue, then
$H$ has a discrete eigenvalue in $(-M,M)$.
By the min-max principle, we need to f\/ind a unit
vector $\Psi$ such that $
\langle \Psi, L(F)\Psi\rangle <0$.
Indeed, for each $f \in D(\Delta)$,
we can f\/ind vectors $\Psi_f^{\pm}\in D(L(F))$,
 such that
$\langle \Psi_f^{\pm},L(F)\Psi_f^{\pm}\rangle=\langle f,S_{\pm}f\rangle$.
By the present assumption, there exists a non-zero vector
$f_0\in D(\Delta)$ such that $\langle f_0, S_+(F)f_0\rangle <0$ or
$\langle f_0, S_-(F)f_0
\rangle <0$. Thus the desired results follow.
\end{proof}

To f\/ind a class of $F$ such that
$E_0(S_+(F))<0$ or $E_0(S_-(F))<0$, we proceed as follows.
For a constant $\varepsilon >0$ and a function $f$ on ${\mathbb R}^d$,
we def\/ine a function $f_{\varepsilon}$ on ${\mathbb R}^d$ by
\[
f_{\varepsilon}(x):=f(\varepsilon x), \qquad x \in {\mathbb R}^d.
\]
The following are key Lemmas.

\begin{lemma}\label{S-lem} Let $V:{\mathbb R}^d\to {\mathbb R}$ be in $L^2_{\rm loc}({\mathbb R}^d)$
and
\[
S_{\varepsilon}:=-\Delta+V_{\varepsilon}.
\]
Suppose that:
\begin{enumerate}\itemsep=0pt
\item[\rm (i)] For all $\varepsilon>0$, $S_{\varepsilon}$
is self-adjoint, bounded below and $\sigma_{\rm ess}
(S_{\varepsilon})\subset [0,\infty)$.
\item[\rm (ii)] There exists a nonempty open set $\Omega\subset
\{x\in {\mathbb R}^d
|V(x)<0\}$.
\end{enumerate}
Then then there exists a constant $\varepsilon_0>0$
such that, for all $\varepsilon\in (0,\varepsilon_0)$,
$S_{\varepsilon}$ has a discrete ground state.
\end{lemma}

\begin{proof} A basic idea for the proof of this lemma is to use
the min-max principle (see~\cite[Lem\-ma~4.3]{AHS}).
\end{proof}

\begin{lemma}\label{S-lem1} $V:{\mathbb R}^d\to {\mathbb R}$ be continuous with
$V(x)\to 0 (|x|\to\infty)$.
Suppose that $\{x\in {\mathbb R}^d|
V(x)<0\}\not=\varnothing$. Then:
\begin{enumerate}\itemsep=0pt
\item[\rm (i)] $-\Delta+V$ is self-adjoint and bounded below.
\item[\rm (ii)] $\sigma_{\rm ess}(-\Delta+V)=[0,\infty)$.
\item[\rm (iii)] $S_{\varepsilon}$ has a discrete ground state
for all $\varepsilon\in (0,\varepsilon_0)$ with
some $\varepsilon_0 >0$.
\end{enumerate}
\end{lemma}

\begin{proof} The facts (i) and (ii) follow from
the standard theory of Schr\"odinger operators.
Part~(iii)~follow from a simple application of Lemma~\ref{S-lem}
(for more details, see the proof of~\cite[Lemma~4.4]{AHS}).
\end{proof}

We now consider a one-parameter family of
Dirac operators:
\[
H_{\varepsilon}
:=(-i){\boldsymbol \alpha} \cdot \nabla+\frac 1{\varepsilon}
M(\beta \otimes I)U_{F_{\varepsilon}}.
\]

\begin{theorem}\label{ex-gs1}
Let $\dim {\cal K}<\infty$. Assume Hypothesis {\rm (A)} and \eqref{F0}.
Suppose that $D_3\cos F$ is not identically zero.
Then there exists a constant $\varepsilon_0>0$
such that, for all $\varepsilon \in (0,\varepsilon_0)$,
$H_{\varepsilon}$ has
a~discrete positive energy ground state or
a~discrete negative ground state.
\end{theorem}

\begin{proof} This follows from Theorem
\ref{ex-gs} and Lemma \ref{S-lem1}
(for more details, see the proof of \cite[Theo\-rem~4.5]{AHS}).
\end{proof}

\section[Symmetry reduction of $H$]{Symmetry reduction of $\boldsymbol{H}$}

Let $T_1$, $T_2$ and $T_3$ be bounded self-adjoint operators on ${\cal K}$ satisfying
\begin{gather*}
 T_j^2=I, \qquad j=1,2,3,\\
T_1T_2=iT_3, \qquad T_2T_3=iT_1, \qquad T_3T_1=iT_2.
\end{gather*}
Then it is easy to see that
the anticommutation relations
\[
\{T_j,T_k\}=2\delta_{jk}I, \qquad j,k=1,2,3
\]
hold. Since each $T_j$ is a unitary self-adjoint operator with $T_j\not=\pm I$,
it follows that
\[
\sigma(T_j)=\sigma_{\rm p}(T_j)=\{\pm 1\}.
\]
We set ${\boldsymbol T}=(T_1,T_2,T_3)$.

In this section we consider the  case where
$T({\boldsymbol x})$ is of the following form:
\[
T({\boldsymbol x})={\boldsymbol n}({\boldsymbol x})\cdot {\boldsymbol T},
\]
where ${\boldsymbol n}({\boldsymbol x})$ is the vector f\/ield in Example \ref{ex1}.
We use the cylindrical coordinates for points ${{\boldsymbol x}}=(x_1,x_2,x_3)\in
{\mathbb R}^3$:
\[
x_1=r\cos\theta, \qquad x_2=r\sin\theta, \qquad x_3=z,
\]
where $\theta \in [0,2\pi),\, r>0$.
We assume the following:

\medskip

\noindent
{\bf Hypothesis (B).} There exists a continuously dif\/ferentiable
function $G:(0,\infty)\times {\mathbb R} \to {\mathbb R}$ such that
\begin{enumerate}\vspace{-2mm}\itemsep=0pt
\item[(i)] $F({{\boldsymbol x}})=G(r,z)$, ${{\boldsymbol x}} \in
{\mathbb R}^3\setminus\{0\}$;
\item[(ii)] $\lim\limits_{r+|z| \to \infty}G(r,z)=0$;
\item[(iii)]
$\sup\limits_{r>0, z\in {\mathbb R}}
(|\partial G(r,z)/\partial r|+|\partial G(r,z)/\partial z|) < \infty$.
\end{enumerate}

We take the vector f\/ield ${\boldsymbol n}:{\mathbb R}^3 \to {\mathbb R}^3$
to be of the form
\[
{{\boldsymbol n}}({{\boldsymbol x}}):=\big(\sin\Theta(r,z)\cos (m\theta),
\sin\Theta(r,z)\sin (m\theta),  \cos\Theta(r,z)\big),
\]
where $\Theta:(0,\infty) \times {\mathbb R} \to {\mathbb R}$ is
continuous and
$m$ is a natural number.

Let $L_3$ be the third component  of the angular
momentum  acting in $L^2({\mathbb R}^3)$ and
\begin{gather}
K_3:=L_3\otimes I +\frac 12 \Sigma_3\otimes I
+\frac {m}{2} I \otimes T_3
\end{gather}
with $\Sigma_3:=\sigma_3\oplus \sigma_3$.
It is easy to see that $K_3$
is a self-adjoint operator acting in ${\cal H}$.

\begin{lemma}\label{com1}
Assume  that
\begin{gather}
\Theta(\varepsilon r,\varepsilon z)=
\Theta(r, z), \qquad
(r,z) \in (0,\infty)\times
{\mathbb R},\qquad \varepsilon>0. \label{Theta}
\end{gather}
Then, for all $t \in {\mathbb R}$ and $\varepsilon>0$,
the operator equality
\begin{gather}
e^{itK_3}H_{\varepsilon}e^{-itK_3}=H_{\varepsilon}  \label{uniteq}
\end{gather}
holds.
\end{lemma}

\begin{proof} Similar to the proof of  \cite[Lemma 5.2]{AHS}. We remark that,
in the calculation of
\[
e^{itK_3}T({\boldsymbol x})e^{-itK_3}
=\sum_{j=1}^3e^{itL_3}n_j({\boldsymbol x})e^{-itL_3} e^{itmT_3}T_je^{-itmT_3},
\]
the following formulas are used:
\begin{gather*}
(T_1\cos mt-T_2\sin mt)e^{itmT_3}=T_1,
\qquad (T_1\sin mt +T_2\cos mt)e^{itmT_3}=T_2.\tag*{\qed}
\end{gather*}
\renewcommand{\qed}{}
\end{proof}

\begin{definition}{\rm
We say that two self-adjoint operators on a Hilbert space
strongly commute if their spectral measures commute.}
\end{definition}

\begin{lemma}\label{stcom}
Assume \eqref{Theta}.
Then, for all $\varepsilon>0$,
$H_{\varepsilon}$ and $K_3$ strongly commute.
\end{lemma}

\begin{proof} By  (\ref{uniteq}) and the functional calculus, we have
for all $s,t\in {\mathbb R}$
$e^{itK_3}e^{isH_{\varepsilon}}e^{-itK_3}=e^{isH_{\varepsilon}}$, which
is equivalent to
$e^{itK_3}e^{isH_{\varepsilon}}=e^{isH_{\varepsilon}}e^{itK_3}$,
$s,t \in {\mathbb R}$. By a general theorem (e.g.,
\cite[Theorem~VIII.13]{RS1}), this implies the strong commutativity
of $K_3$ and $H_{\varepsilon}$.
\end{proof}

Lemma \ref{stcom} implies that $H_{\varepsilon}$ is reduced
by eigenspaces of $K_3$.
Note that
\begin{gather*}
\sigma(K_3) =\sigma_{\rm p}(K_3) =\left\{
\ell+\frac {s}2+\frac {mt}{2}
\, \bigg|\, \ell \in {\mathbb Z}, s=\pm 1, t=\pm 1\right\}.
\end{gather*}
The eigenspace
of $K_3$ with eigenvalue $\ell+({s}/2)+ ({mt}/{2})$
is given by
\[
{\cal M}_{\ell,s,t}:={\cal M}_{\ell}\otimes {\cal C}_s\otimes
{\cal T}_t
\]
with ${\cal C}_s:=\ker(\Sigma_3-s)$ and
${\cal T}_t:=\ker(T_3-t)$.
Then ${\cal H}$ has the orthogonal decomposition
\[
{\cal H}=\oplus_{\ell \in {\mathbb Z}, s,t \in \{\pm 1\}}
{\cal M}_{\ell,s,t}.
\]
Thus we have:

\begin{lemma}
Assume \eqref{Theta}.
Then, for all $\varepsilon>0$,
$H_{\varepsilon}$
is reduced by each ${\cal M}_{\ell,s,t}$.
\end{lemma}

We denote by
$H_{\varepsilon}(\ell,s,t)$  by the reduced part of
$H_{\varepsilon}$ to ${\cal M}_{\ell,s,t}$ and
set
\[
H(\ell,s,t):=H_1(\ell,s,t).
\]

For $s=\pm 1$ and $\ell \in {\mathbb Z}$, we def\/ine
\[
L_{s}(G,\ell):=
-\frac {\partial^2}{\partial r^2}
-\frac 1r \frac{\partial}{\partial r}
+\frac {\ell^2}{r^2}+\frac{\partial^2}{\partial z^2}
+s M D_z\cos G
\]
acting in $L^2((0,\infty)\times {\mathbb R},rdrdz)$
with domain
\[
D(L_{s}(G,\ell))
:=C_0^{\infty}((0,\infty)\times {\mathbb R})
\]
and set
\[
{\cal E}_0(L_{s}(G,\ell)):=
\inf_{f\in C_0^{\infty}((0,\infty)\times {\mathbb R}),
\|f\|_{L^2((0,\infty)\times {\mathbb R},rdrdz)}=1}
\langle f,L_{s}(G,\ell)f\rangle.
\]

The following theorem is concerned with
the existence of discrete ground states of $H(\ell,s,t)$.

\begin{theorem}\label{ex-gs2} Assume Hypothesis {\rm (B)} and \eqref{Theta}.
Fix an $\ell\in {\mathbb Z}$  arbitrarily, $s=\pm 1$ and $t=\pm 1$.
Suppose that  $\dim {\cal T}_t <\infty$ and
\[
{\cal E}_0(L_{s}(G,\ell)) < 0.
\]
Then $H(\ell,s,t)$ has
a discrete positive energy ground state or
a discrete negative ground state.
\end{theorem}

\begin{proof} Similar to the proof of Theorem \ref{ex-gs}
(for more details, see the proof of \cite[Theo\-rem~5.5]{AHS}).$\!\!$\end{proof}

\begin{theorem}\label{ex-gs3}
Assume Hypothesis {\rm (B)} and \eqref{Theta}.
Suppose that $\dim{\cal T}_t<\infty$ and that $D_z \cos G$
is not identically zero.
Then, for each $\ell\in {\mathbb Z}$,
there exists a constant $\varepsilon_{\ell}>0$
such that, for all $\varepsilon \in (0,\varepsilon_{\ell})$,
each $H_{\varepsilon}(\ell,s,t)$ has
a discrete positive energy ground state or
a discrete negative ground state.
\end{theorem}

\begin{proof} Similar to the proof of Theorem \ref{ex-gs1}
(for more details, see the proof of \cite[Theo\-rem~5.6]{AHS}).$\!\!$
\end{proof}

Theorem \ref{ex-gs3} immediately  yields the following result:

\begin{corollary}
Assume Hypothesis {\rm (B)} and \eqref{Theta}.
Suppose that $\dim{\cal T}_t<\infty$ and that $D_z\cos G$
is not identically zero. Let $\varepsilon_{\ell}$ be as in
Theorem~{\rm \ref{ex-gs3}} and, for each $n\in {\mathbb N}$ and $k>n$
($k,n\in {\mathbb Z}$),
$\nu_{k,n}:=\min\limits_{n+1 \leq  \ell\leq k}\varepsilon_{\ell}$.
Then, for each $\varepsilon \in (0,\nu_{k,n})$,
$H_{\varepsilon}$ has at least $(k-n)$ discrete eigenvalues counting
multiplicities.
\end{corollary}

\begin{proof} Note that $\sigma_{\rm p}(H_{\varepsilon})=\cup_{\ell \in {\mathbb Z},
s,t=\pm 1}\sigma_{\rm p}(H_{\varepsilon}(\ell,s,t))$.
\end{proof}

\section{A unitary transformation}

We go back again to the generalized CQSM def\/ined in Section~2. It is easy to see that
the operator
\[
X_F:=\frac {1+\gamma_5}2\exp\left(
iF\otimes \frac{T}{2}\right)+\frac {1-\gamma_5}2\exp\left(-iF\otimes
\frac{T}{2}\right)
\]
is unitary.
Under Hypothesis (A), we can def\/ine the following operator-valued
functions:
\[
B_j({{\boldsymbol x}}):=\frac 12
D_j[F({{\boldsymbol x}})T({\boldsymbol x})], \qquad {{\boldsymbol x}}\in {\mathbb R}^3, \qquad j=1,2,3.
\]
We set
\[
{\boldsymbol B}:=(B_1,B_2,B_3)
\]
and introduce
\[
H({\boldsymbol B}):=(-i){\boldsymbol \alpha}\cdot\nabla+M\beta -
{\boldsymbol \sigma} \cdot {\boldsymbol B}
\]
acting in ${\cal H}$.
Since ${\boldsymbol \sigma}\cdot {\boldsymbol B}
$ is a bounded self-adjoint operator,
$H({\boldsymbol B})$ is self-adjoint with $D(H({{\boldsymbol B}}))=\cap_{j=1}^3
D(D_j\otimes I)$.

\begin{proposition}\label{HB} Assume Hypothesis {\rm (A)} and that
$T({\boldsymbol x})$ is independent of ${\boldsymbol x}$.
Then
\[
X_F H X_F^{-1}=H({{\boldsymbol B}}).
\]
\end{proposition}

\begin{proof} Similar to the proof of \cite[Proposition 6.1]{AHS}.
\end{proof}

Using this proposition, we can prove the following theorem:

\begin{theorem} Let $\dim {\cal K} <\infty$. Assume Hypothesis {\rm (A)} and that
$T({\boldsymbol x})$ is independent of ${\boldsymbol x}$.
Suppose that
\[
\lim_{|{\boldsymbol x}|\to\infty}|\nabla F({\boldsymbol x})|=0.
\]
Then
\begin{gather}
\sigma_{\rm ess}(H)=(-\infty,-M]\cup [M,\infty).\label{espec2}
\end{gather}
\end{theorem}

\begin{proof} By Proposition~\ref{HB}, we have
$\sigma_{\rm ess}(H)=\sigma_{\rm ess}(H({\boldsymbol B}))$. By the present assumption,
$B_j({\boldsymbol x})=D_jF({\boldsymbol x}) T({\boldsymbol 0})/2$. Hence
\[
\sup_{|{\boldsymbol x}|>R}\|{\boldsymbol \sigma} \cdot {\boldsymbol B}({\boldsymbol x})\|
\leq \sum_{j=1}^3(\|T({\boldsymbol 0})\|/2)\sup_{|{\boldsymbol x}|>R}|D_jF({\boldsymbol x})|
 \to 0 \qquad (R\to\infty).
\]
Therefore, as in the proof of Theorem \ref{espec}, we conclude that
$\sigma_{\rm ess}(H({\boldsymbol B}))=(-\infty,-M]\cup [M,\infty]$.
Thus (\ref{espec2}) follows.
\end{proof}

\subsection*{Acknowledgements}

The author would like to thank N.~Sawado for
kindly informing on typical examples of
prof\/ile functions and comments.
This work was supported by the Grant-In-Aid 17340032 for
Scientif\/ic Research from the JSPS.

\LastPageEnding

\end{document}